\documentclass[epsfig,graphics,twocolumn,floatfix,a4paper,showpacs]{revtex4}
\usepackage{graphicx}
\usepackage{dcolumn}
\usepackage{bm}

\begin{document}
%
%
%

\title{Symmetry breaking in small rotating cloud of trapped ultracold
Bose atoms}
\author{D. Dagnino$^1$, N. Barber\'an$^1$, M. Lewenstein$^{2,3}$, K.
Osterloh$^3$, and A. Riera$^1$}
\affiliation{$^1$Dept. ECM,
Facultat de F\'\i sica, U. de Barcelona,
E-08028 Barcelona, Spain}
\affiliation{$^2$ICREA and ICFO--Institut de Ci\`encies Fot\`oniques,
Av. del
Canal Ol{\'\i}mpico s/n, 08860 Castelldefells, Barcelona, Spain}
\affiliation{$^3$Institute for Theoretical Physics, University of
Hannover, Appelstrasse 2, 30167 Hannover, Germany}

\begin{abstract}
We study the signatures of rotational and phase symmetry breaking in
small rotating
clouds of trapped ultracold Bose atoms by looking at rigorously
defined
condensate wave function. Rotational
symmetry
breaking occurs in narrow frequency windows, where the ground
state of
the system has  degenerated with respect to the total angular momentum,
and
it leads to a complex wave function that exhibits vortices clearly
seen as holes in the density, as well as characteristic vorticity.
Phase symmetry (or gauge symmetry) breaking, on
the other hand, is clearly manifested
in the interference of two independent rotating
clouds.
\end{abstract}

\pacs{73.43.-f,05.30.Jp, 03.75.Hh}

\maketitle

Symmetry breaking in finite systems has been a subject of intensive
debate in
physics, in general (cf. the Ref.\cite{blaizot}), and in physics of
ultracold gases
in particular over the years. For Bose-Einstein condensates (BEC)
two symmetries
play a particular role: $U(1)$ phase symmetry and $SU(2)$ (or $SO(3)$)
rotational symmetry. In the large $N$ limit, one breaks these
symmetries by
hand, as proposed originally by Bogoliubov \cite {bogo}. Thus,
the accurate
way to deal with macroscopic Bose Einstein condensates (BEC's)
is by the use of a classical field, also called an order parameter, or
the wave function (WF) of the condensate. This function is a single
particle
(SP) wave function, which is the solution of the
Gross Pitaevskii (GP) equation within the mean field approximation,
that characterizes the system in a proper
way \cite{but}. It has an arbitrary, but fixed phase, and for
rotating systems
with more than one vortex it exhibits arbitrarily places, but fixed
vortex array. For dilute ultracold Bose gases (i.e. when $n|a|^3<<1$
\cite{pit}
where $n$ is the density and $a$ is the s-wave scattering length)
mean field,
or Bogoliubov approach  is capable to reproduce very well the main
properties,
despite the fact for finite, fixed $N$ and total angular momentum $L$,
which are both constants of the motions,  mean field  theory cannot
be exact.
This observation has stimulated a lot of discussion about the nature
of the
phase of BEC \cite{phase1,phase2}, and particle-conserving Bogoliubov
approach \cite{particle}.
The modern point of view (for a recent discussion see \cite{mul})
implies that two BEC's with fixed $N$ each one, will produce a well
defined interference
pattern of fringes as a result of the measurement in only one shot
(comparable with the calculated n-correlation function) in contrast
with the density, which would be obtained as a mean image of random
interference patterns from several shots.
The position of fringes in
the given measurement are determined by subsequent localization of
atoms arriving
at detectors; the first atom is completely random, second is
correlated,
third even more correlated etc. \cite{phase2}. Thus the information
about
the pattern is obtained from the many-body wave function by looking
at pair,
triple, ... correlations. The breaking of rotational symmetry
should occur in large
rotating clouds  in the similar way, and a pure L-state would
show, in a time-of-flight experiment, a definite interference pattern
accurately represented by n-correlation functions, different from a
circular symmetric profile of the single particle density. It would be
a test of the meaning assigned to the measurement. Unfortunately, for
large N-systems, the total angular momentum of the stationary states
is not well defined and there is no qualitative difference between
density and n-correlation function, usually showing in both cases
vortex arrays. For small rotating
clouds the
situation is, however, different, as we have shown in Ref. \cite{bar}.
Typically, the GS's are pure-L states for most of the values of
$\Omega$.
Only,
in the very narrow window of frequencies, where the ground states is
degenerated with respect to $L$, vortex arrays can be obtained,
arbitrary small symmetry breaking deformation
of the trap potential leads to the appearance of
symmetry breaking vortex arrays both in density and pair-correlations.
Namely, in the regime of pure L-GS small systems would provide a
suitable test for the meaning of the measurement distinguishing
between the density or the pair-correlation output.

In this Letter we study the effects of symmetry breaking in small
rotating clouds
of trapped ultracold Bose atoms in more depth, by introducing the
rigorous
definition of the condensate wave function, defined as an eigenvector
of
the one body density matrix operator (OBDM), corresponding to
the largest eigenvalue.
Such definition of the order parameter has been introduced in
classical
papers on {\it off-diagonal long range order} \cite{class}. It has,
however,
rarely been used since its application requires the knowledge of the
full many-body wave function, or at least of the exact OBDM.
Since for  quantum gases exact analytic solutions are either not known
(2D and 3D), or very difficult to handle (1D), so far this definition
has been only applied to the case of model system with harmonic forces.
Here we apply
for the first time to the rotating gas, using exact numerically
calculated OBDM for few atom systems. We identify in this way possible
states with vortices, and
obtain phase characteristics of the wave function (reflecting
quantized circulation of
vortices), and provide unambiguous definition of the degree of
condensation.
With such calculated order parameter we then reproduce the density
and interference patterns for two
condensed clouds, and shed new light on the discussion of the origins
of symmetry
breaking in finite mesoscopic systems.

We consider a two-dimensional system of few
Bose atoms trapped in a parabolic rotating trap around the z-axis.
The rotating frequency $\Omega$ is strong enough to consider the
Lowest Landau
Level regime with atoms interacting via  contact forces.
Our main goal
is the description of the stationary states for different values of
$\Omega$, analyzed from the rotating frame of reference, unless
otherwise stated. Our analysis is  performed using  the exact
diagonalization
formalism, valid for arbitrary interactions and densities.
However, in contrast to the mean field approach, this
method deals
with multi-particle WF's and loses the intuitive picture
provided by the mean field order parameter. Our goal is to obtain in
a precise way a complex scalar field that models efficiently
the system,  and
allows to reproduce the important features, such as  the vortex states.
In
the regime of relatively low
rotation frequency, where the degree of condensation is high and some
vortices appear distributed in an ordered arrays,
this scalar field plays the role of a genuine order parameter.
On the other
hand, it looses its capability to represent the system as
$\Omega$
approaches the melting point, where the prediction \cite{coo} is
that the
vortex lattice disappears and the systems turns, for large systems,
into a Laughlin liquid.

The way to know if there is a "macro-occupied"  SP wave
function in the ground state $|GS\rangle$ is to
look at the eigenvalues and eigenvectors of the OBDM
\cite{pit,class}.  That is to say, one must solve the eigenvalue
equation
\begin{equation}
\int d\vec{r'} n^{(1)}(\vec{r},\vec{r'})
\psi^{*}_l(\vec{r'})= n_l
\psi^{*}_l(\vec{r}),
\end{equation}
where
\begin{equation}
n^{(1)}(\vec{r},\vec{r'}) = \langle GS\mid \hat{\Psi}^{\dag}(\vec{r})
\hat{\Psi}(\vec{r'})|GS \rangle,
\end{equation}
with $\hat{\Psi}$ being the field operator. If there exist a relevant
eigenvalue $n_1\gg n_l$ for $l=2,3,\ldots$, then
\begin{equation}
\sqrt{n_1} \psi_1(\vec{r})e^{i\phi_1}
\end{equation}
plays the role of the order parameter of the system, where $\phi_1$
is an arbitrary constant phase.
The WF may be expanded in the form
$
\psi_1(\vec{r})=\sum_{l=0}^m \beta_{1l}\varphi_l(\vec{r})$,
using the complete set of Fock-Darwin \cite{jac} WF's given
by
$
\varphi_l(\vec{r})=e^{il\theta}r^le^{-r^2/2}/\sqrt{\pi l!}
$,
where $l$ labels the single particle  angular momentum, and $m$ is
equal to the largest total angular momentum $L$ involved in the
expansion of the GS; length unit is here
$\lambda=\sqrt{\hbar/(m\omega_{\perp})}$, and
$\omega_{\perp}$ denotes the trap frequency.
The same SP basis is used in our numerical simulations to represent
both the
 field
operator and the multiparticle GS wave function.

An alternative, and perhaps even more appropriate SP basis is
determined by
the functions $\psi_l(\vec{r})$. One can define a set of canonical
creation
 and annihilation operators for them:
\begin{equation}
\hat a^{\dag}_l=\int d\vec{r'} \psi^{*}_l(\vec{r'})\hat{\Psi}(
\vec{r'}),
\end{equation}
and $\hat{a}_l$ being the hermitian conjugate of $ \hat a^{\dag}_l $.
The Hilbert
space attain then a tensor structure with respect to the modes
$\hat a_l$, and the
new Fock (occupation number) many body basis $|n_1\rangle\otimes |n_2
\rangle \otimes \ldots$. The macro-occupied mode contains on average
$n_1$ atoms, but
this number fluctuates, and most presumably normally, i.e. the
fluctuations
of $n_1$ are of order $\sqrt{n_1}$; to this aim one has to calculate
$\langle GS|(\hat a^{\dag}_1 \hat a_1)^2|GS\rangle$. This implies
that atom number
fluctuations between the macro-occupied mode (condensate) and the rest
of the modes
(that could be regarded as phonon modes, quasi-particles) will tend to
reduce the fluctuations of the phase. A natural consequence of this
observation is to expect that a very fine approximation of the GS is
given
 by the coherent state $|
\alpha_1 \rangle$, such that $ \hat a_1$ $|
\alpha_1 \rangle =\sqrt{n_1} \psi_1 e^{i\phi_1}$ $|
\alpha_1 \rangle $. If $n_l$ for $l=2,3,\ldots$ are very small
we may neglect them, and
approximate the many body wave function by $|\alpha_1\rangle\otimes
|0_2\rangle\otimes|0_3\rangle\otimes\ldots$. In a more precise
description,
we should rather approximate the GS by
$|\alpha_1\rangle\otimes|\alpha_2
\rangle\otimes|\alpha_3\rangle\otimes\ldots$,
where $\hat a_l |\alpha_l\rangle= \sqrt{n_l} \psi_l
e^{i\phi_l}$ $|
\alpha_l \rangle$ where the phases $\phi_l$ are arbitrary; one
should, however, choose
them to be random in order to reproduce (on average) the same OBDM
as the one obtained by exact numerical diagonalization.

This representation implies that the next
simplifiying step would be the representation  of the GS by a clasical
field
entering the GP equation, and
containing all the involved coherent states $| \alpha_k\rangle$,
$k=1,...m+1$ as,
$
\Psi(\vec{r})=\sum_{k=1}^{m+1} \sqrt{n_k} \psi_k e^{i\phi_k}\,\,\,
$ with random phases. Calculation of quantum mechanical averages
would then in principle
require averaging over random phases, which makes this approach
technically difficult.

As long as the exact GS is a state with well defined angular momentum,
(a pure $L$-state) not degenerated with other lowest energy states in
different L-subspaces,
it is easy to demonstrate that the FD functions are the eigenstates
of
Eq.(1) and the eigenvalues $n_l$ are the occupations usually used in
literature. However, at certain values of $\Omega$ where degeneracy
takes
place and vortex states without circular symmetry (except the case
of only one centered vortex) are possible \cite{bar},
the eigenfunctions of Eq.(1) are linear combinations of the FD
functions and the macro-occupied function $\psi_1$
that represents the vortex
state has expected SP angular momentum
given by
$
\hbar \tilde{l}=\sum_j \mid \beta_{1j}\mid^2 \hbar l_j
$
where $l_j$ are integers.

A convenient definition of the degree of condensation which senses
the loss of macro-occupation is given by
\begin {equation}
c=\frac{n_1-\tilde{n}}{N}
\end {equation}
where $N$ is the total number of atoms and $\tilde{n}$ is the mean
occupation calculated without the first value $n_1$.

In what follows, we show some results that confirm the convenience to
represent the whole system by $\psi_1$ at certain values of $\Omega$.
As a general result, for vortex states, $n_1$ is always larger than
the occupation of the most important FD state within the exact GS. In
addition, $\psi_1$ provides a non-ambiguous way to
characterize
vortices, not only showing dimples in the density profile, but also
indicating the position of each one by the change on multiples of
$2\pi$ of the phase $S(\vec{r})$ in $\psi_1(\vec{r})=\mid
\psi_1(\vec{r})\mid e^{iS(\vec{r})}$ when moving around each
vortex. In
Fig.1 for $N=6$ atoms, we show for three different values of $\Omega$
where degeneracy takes place, the comparison between the contour plots
of the density of the exact GS and the density of $\psi_1$, as well as
the map of the phase $S(\vec{r})$ of $\psi_1$. In the first case (a)
the GS contains
two vortices that appear in a clearer way in the order parameter, as
it excludes the non condensed part that smears the structure of the
GS. The same picture is shown in (b) where
four vortices become visible. In the second case,
the map of the phase not only localizes vortices with one unit of
quantized circulation, but also indicates that incipient vortices are
growing at the edge of the system.
In the last case (c),
a six-fold symmetry is obtained not attached in this case to
vortices, but to a mixed structure of dimples and bumps, a precursor
of the Wigner type structure observed for few atoms in the Laughlin
state at an angular momentum of $L=30$ \cite{bar}.
The degree of condensation as defined in Eq.(5)
decreases as $0.343$, $0.192$ and $0.015$ from (a) to (c).
The order in vortices
and disorder in atoms evolves to order in atoms.
As $\Omega$ approaches the frequency of the trap, the
occupations tend to equalize and in the Laughlin state, where $n_l$
are the FD occupations  (since it is a pure L-state), and the
degree of condensation tends to zero.
\begin{figure}[htb]
\includegraphics*[width=1\columnwidth]{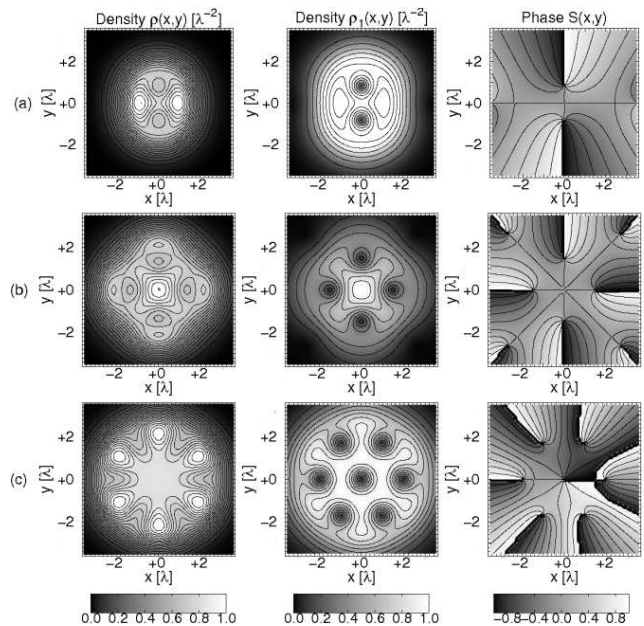}
\caption{
 For $N=6$ the first two pictures on each row show the density
contour plot
of the GS ($\rho(x,y)$) and the $\psi_1$ function
($\rho_1(x,y)$) respectively. The third picture
shows tha map of the phase $S(\vec{r})$ (see text). (a) shows a two
vortex
structure at $\Omega=0.941$ (where degeneracy between $L=10$ and
$12$ takes place).
(b) shows a four vortex structure, $\Omega=0.979$ (degeneracy between
$L=20,22$ and
$24$). (c) shows a six-fold structure, $\Omega=0.983$ (degeneracy
betwee
$L=24,26,28$ and $30$). In all cases $\omega_{\perp}=g=1$ in units of
$\lambda$ and $u=\hbar \omega_{\perp}$.}
\end{figure}

Some excited states with large L can also be analyzed.
For $N=3$
and $L=9$ we obtain a large vortex state with three
units of circulation. Such state has been predicted  in previous
theoretical
studies as a possible  giant
vortex GS (with all vorticity  confined to the center of the
condensate), in the presence of  a small quartic potential  added
to the parabolic
trap. In such a case
stationary states for  $\Omega > \omega_{\perp}$ are possible
\cite{dan}. In our calculations the giant vortex appears as an excited
state, anticipating
this possibility. So far there is no
experimental evidence
of giant vortex structures in bosonic systems \cite{bre}, they
have been reported in superconductive disks \cite{kan}.

Finally we show the interference pattern produced by the overlap of
two initially independent condensates represented by $\psi_1$
functions. This study is motivated by
an increasing amount of recent work  revealing  the possibility
of  obtaining very detailed  experimental information on the
interference
pattern produced not only during the overlap of two, or more
independent
condensates \cite{and}, but also within a unique
condensate \cite{rit}.

The idea underlying our assumption is the following: we
represent the two independent condensates which we call $a$ and $b$
by their macroscopic occupied function $\psi_a$ and $\psi_b$
respectively. By this we mean that the condensates are in two unknown
coherent
states $|\alpha_a\rangle$ and $|\alpha_b \rangle$ from which
we know their order parameter
except for their constant phases $\phi_a$ and $\phi_b$ (see Eq.(3)).
At time
$t=0$ s the condensates are separated by a
distance $d$ and the traps are switched off.
The
time evolution of the system is obtained (once the transformation
to
the laboratory frame of reference is performed, multiplying the
functions
by  $exp(-i\Omega t \hat{L_z})$) in three steps: First, the Fourier
transform
of the total order parameter (the sum of the two contributions)
is performed.
Then, the time
evolution of the Fourier components by multiplying them by
exponentials of the type
$exp(i\hbar k^2 t /2m)$ is realised; this step is done under
the assumption
that during the time-of-flight
the interactions are irrelevant.  Finally, in the third step,
inverse
Fourier transformation is performed. The results are shown in Fig.2
where three
different times are considered. Fortunately, the uncertain about the
phase relation $\phi=\phi_a-\phi_b$ is not important in the case
considered, as only two
terms are involved and a change on the relative phase would
only produce a global shift of the interference pattern.
\begin{figure}[htb]
\includegraphics*[width=0.8\columnwidth]{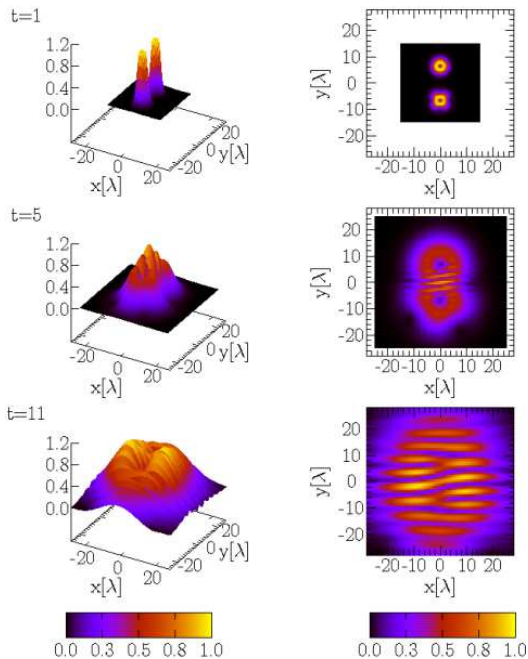}
\caption{
Time evolution of the interference pattern during the overlap of
two released condensates initially separated by a distance
$d=15\lambda$.
Initially each condensate contains $N=6$ atoms and their GS are
characterized by $L=6$ at $\Omega=0.019$ and by a mixture of $L=6,8$
and $10$ at $\Omega=0.0847$ respectively (all quantities are in units
of $\lambda$ and $u$).}
\end{figure}

We conclude that, we have demonstrated that the use of the
eigenfunctions of the OBDM operator provides a
useful and precise tool to analyze the exact GS obtained from exact
diagonalization and specially the vortex states. These eigenfunctions
localize and quantize the vortices and reproduce the time evolution
of the interference pattern of two overlapping condensates. We want to
point out that our results imply an alternative interpretation about a
subject that has attracted much attention recently related with the
interference pattern formation. One possibility suggested by Mullin
and
collaborators \cite{mul} is that the experimental
measurement projects the initial
condensates in  Fock states into phase states, the atom distribution
between the two components become uncertain and the pattern formation
is possible. The other possibility discussed by Cederbaum et
al. \cite{ced},
is that the interference pattern appears if one includes interaction
during the time-of-flight even for states that initially are Fock
states.
In our case, the real initial states are Fock states and  no
interaction
is included during the time-of-flight. However, we assume that
the degree of
condensation of the initial states is large enough to
be properly
represented by an order parameter (condensate wave function).
Fluctuations of the number
of condensed atoms reduce the phase fluctuations and determine the
order parameter phase. In effect, exact ground state manifest
themselves as phase states even for small number of particles,
and in this way the
interference patter is produced. Note, however, that in our picture
the process of determination of phase is itself random, and various
phases $\phi_k$
are expected to show up from shot to shot.

We thank J.M. Pons and J. Taron for fruitful discussions. We
acknowledge financial support of MEC Projects
DUQUAG, Consolider QOIT, EUIP SCALA and contracts FIS2004-05639 and
2005SGR00343.


\end{document}